\documentclass[final,5p,times,bibauthor,twocolumn]{elsarticle}
 \usepackage{epsfig}
 \usepackage{graphicx}
 \usepackage{multicol}
 \usepackage{multirow}
 \usepackage{mathrsfs} 
 \usepackage{hyperref}
\journal{Radiation Measurements}
\begin{document}
   
\begin{frontmatter}

\title{Magnetic Monopole Search with the SLIM
  experiment}
\author[label1,label2,label7]{E. Medinaceli}
\author{for the SLIM collaboration$^*$; Presented to the $24^{th}$
  ICNTS.}
\address[label1]{Dip. di Fisica, Universit\'a di Bologna, I-40127 Bologna, 
  Italy}
\address[label2]{INFN Sez. di Bologna, I-40127 Bologna, Italy}
\address[label7]{Laboratorio de F\'{i}sica C\'osmica de Chacaltaya, UMSA, La 
  Paz, Bolivia}
\cortext[collab]{The {\textbf{SLIM}} collaboration:
S. Balestra, S. Cecchini, M. Cozzi, M. Errico, F. Fabbri, G. Giacomelli, 
R. Giacomelli, M. Giorgini, A. Kumar, S. Manzoor, J. McDonald, G. Mandrioli, 
S. Marcellini, A. Margiotta, E. Medinaceli,
L. Patrizii, J. Pinfold, V. Popa, I.E. Qureshi, O. Saavedra,
Z. Sahnoun, G. Sirri, M. Spurio, V. Togo, A. Velarde,
A. Zanini}
\begin{abstract}
The SLIM experiment was an array of 427 $m^2$ of nuclear track detectors, exposed
at a high altitude laboratory (Chacaltaya, Bolivia, 5230 m a.s.l.),
for $\sim$4.22 years. SLIM was sensitive to downgoing intermediate mass magnetic
monopoles with masses of $10^5$ GeV $\leq$ 
$M_M$ $\leq$ $10^{12}$ GeV. The analysis of the full detector gives a
flux upper limit of $1.3\times10^{-15}~cm^{-2}s^{-1}sr^{-1}$ (90\%
C.L.) for downgoing fast intermediate mass magnetic
monopoles. 
\end{abstract}
\end{frontmatter}
\section{Introduction}\label{sec:intro}
P.A.M. Dirac introduced the concept of the magnetic charge "g" in
order to explain the quantization of the electric charge "e". He obtained the
formula $eg=n\hbar c/2$, from which $g=ng_D=n\hbar c/2e=n68.5e$, where
$n=1,~2,~3,~...$ (Dirac, 1931). Magnetic Monopoles (MMs) possesing an
electric charge and bound systems of a MM with an atomic nucleus are
called dyons (Giacomelli, 2000).

The electric charge is naturally quantified in Grand Unified
Theories (GUT) of the strong and electroweak interactions which
descrives phase transitions of the early Universe at the mass 
scale $M_G\sim10^{14}\div 10^{15}$ GeV. These models imply the
existence of \textit{GUT monopoles} with calculable properties such as
their masses. The  MM mass is related to the mass of the X, Y carriers
of the unified interaction, $ m_{M}\ge m_{X}/G$, where G is the
dimensionless unified coupling constant at energies E $\simeq m_{X}$. 
If $m_{X}\simeq 10^{14}-10^{15}$ GeV and $G\simeq0.025$,
$m_{M}>10^{16}-10^{17}$ GeV (Ambrosio, 2002). 

Some GUT models and Supersymmetric models predict \textit{Intermediate
Mass Monopoles} (IMMs) with masses $m_M$ $\sim10^{5} \div 10^{12}$
GeV and magnetic charges of integer multiples of $g_D$. IMMs may have
been produced in later phase transitions in the early Universe and
could be present in the cosmic radiation and may be accelerated up to
relativistic velocities in one coherent galactic magnetic field
domain. Thus one may look for downgoing, fast ($\beta\ge0.03$) heavily
ionizing IMMs (Balestra, 2008; Giacomelli, 2007).

\par
\begin{figure}[htb]
\hspace{-0.5cm}{\centering\resizebox*{8.3cm}{6cm}
{\includegraphics{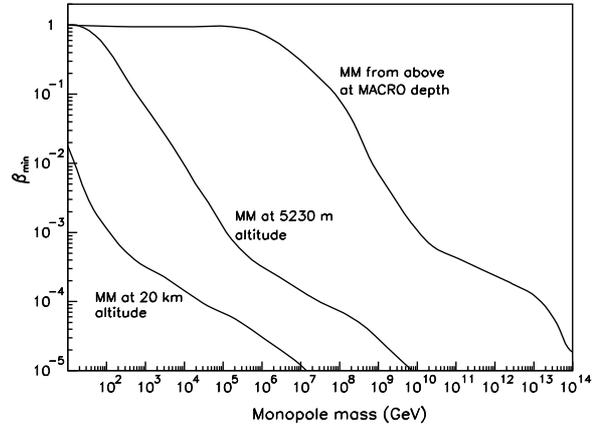}}\par}
\caption{Accessible regions in the plane (mass, $\beta$) for
  monopoles with magnetic charge $g=g_D$ coming from above for
  experiments at altitudes of 20 km, at 5230 m a.s.l.,
  and for an underground detector.}\label{fig1} 
\end{figure}

 The exposure at a high altitude laboratory allows to search for IMMs
of lower masses, higher charges and lower velocities. The main purpose
of the SLIM (Search for LIght magnetic Monopoles) experiment deployed
at the Chacaltaya laboratory in Bolivia at 5230 m a.s.l., was the
search for IMMs (Balestra, 2008; Bakari, 2000).
Fig. \ref{fig1} shows the accessible regions in the plane (mass,
$\beta$) for experiments located at different altitudes. The SLIM
detector was also sensitive to strange quark matter nuggets 
(Witten, 1984) and Q-balls 
(Coleman, 1985); the results on these dark
matter candidates are discused in (Shanoun, 2008).

\section{Experimental procedure}\label{sec:exp}
\begin{figure}[htb]
\vspace{-0.3cm}
 {\centering\resizebox*{4cm}{5cm}{\includegraphics{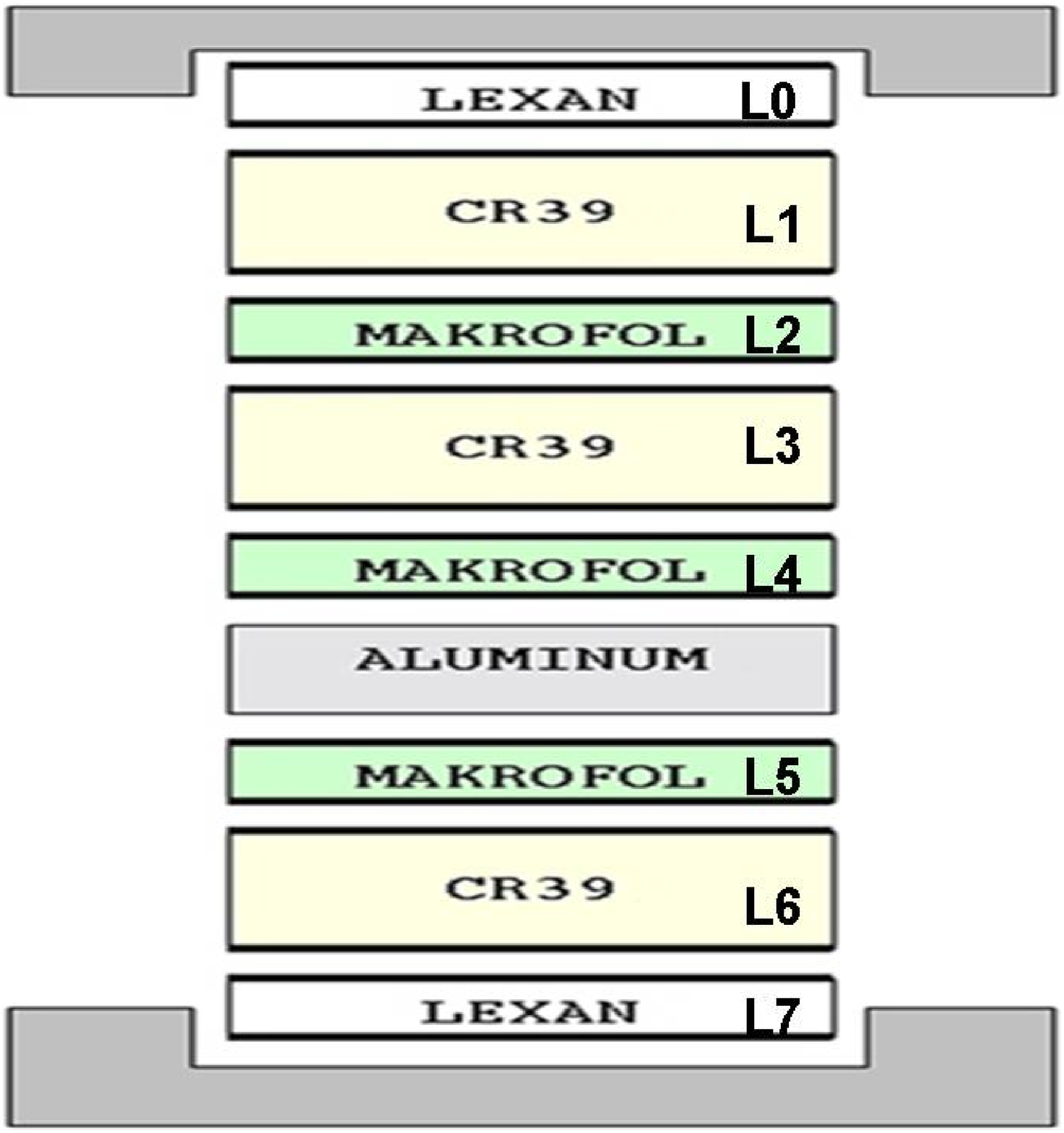}}\par}
\vspace{-0.3cm}
\caption{SLIM module composition.}
\label{fig:stack}
\vspace{-0.2cm}
\end{figure}

The SLIM detection area (427 $m^2$) was an array organized into 7410 
modules, each one of area $24 \times 24$ cm$^2$ (Bakari, 2000).
 All modules were made up of: three layers of CR39$^{\scriptsize
   \textregistered}$~\footnote{The 
  SLIM CR39 was produced  by the Intercast Europe Co, Parma, Italy
  according to our specifications.}, each 1.4 mm thick; 3 layers of
Makrofol DE$^{\scriptsize \textregistered}$~\footnote{Manufactured by
  Bayer AG, Leverkusen, Germany.}, each 0.48 mm thick; 2 layers of
Lexan each 0.25 mm thick and one layer of aluminum absorber 1 mm thick
(see Fig. \ref{fig:stack}). An special batch of 50 m$^2$ of the total
area, was composed of stacks using CR39 containing 0.1\% of DOP
additive. Each stack was sealed in an aluminized plastic bag (125
$\mu$m thick) filled with dry air at a pressure of 1 bar. The stacks
were deployed under the roof of the Chacaltaya Laboratory, roughly 4 m
above ground. 
The geomagnetic cut-off for cosmic rays at the detector latitude is
$\sim12.5~GV$. Environmental conditions at the Chacaltaya lab are:
atmospheric preassure $\sim$0.5 atm; mean temperature $12^{\circ}C$,
oscillating from 0 to 30$^{\circ}C$; the radon concentration is
$\sim$40$\div$50 $Bq/m^3$. The neutron flux measured with BTI bubble 
counters at the detector site, in the range from few hundred
keV to $\sim$20 MeV is $(1.7\pm0.8)\times 10^{-2}~cm^{-2}s^{-1}$
(Zanini, 2001).

{\bf{Etching proceadures and NTD calibrations:}} The passage of a
magnetic monopole in NTDs, such as CR39, 
causes structural line damage in the polymer (forming the so called 
``latent track''). Since IMMs have a constant energy loss 
through the stacks, the subsequent chemical etching should result in
collinear etch-pit cones of equal size on both faces of each detector sheet. 
 In order to increase the  detector ``signal to noise'' ratio different 
etching conditions (Cecchini, 2001; Balestra, 2007; Togo 2008) 
were defined.
 The so-called ``strong etching'' technique produce better surface quality 
and larger post-etched cones. This makes etch pits easier 
to detect under visual scanning. Strong etching was used to analyze the 
top-most CR39 sheet in each module.
``Soft etching''  was applied to the other CR39 layers in a module if a 
candidate track was found after the first scan. 

For CR39 and CR39(DOP) the strong etching conditions were: 8N KOH + 
1.5\% ethyl alcohol at 75~$^\circ$C for 30 hours. The bulk etching velocities 
were $v_B = 7.2\pm0.4~\mu$m/h and $v_B = 5.9\pm0.3~\mu m/h$ for 
CR39 and CR39(DOP), respectively. The soft etching conditions were 6N
NaOH + 1\% ethyl alcohol at 70~$^\circ$C  for 40 hours for CR39 and  
CR39(DOP). The bulk etching rates were $v_B=1.25 
\pm0.02~\mu m/h$ and $v_B = 0.98\pm0.02~\mu m/h$ for CR39 and 
CR39(DOP), respectively. Makrofol NTDs were etched in 6N KOH + 20\%
ethyl alcohol at 50~$^\circ$C for 10 hours; the bulk etch velocity was
$v_B = 3.4~\mu$m/h.\\
{\bf{NTD Calibrations:}}
\begin{figure}[h!]
{\centering\resizebox*{6.5cm}{6.5cm}{\includegraphics{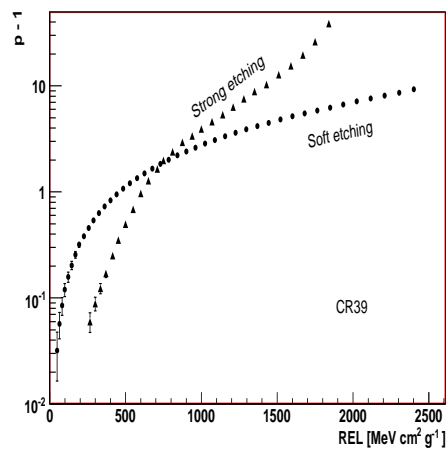}}\par}
\vspace{-0.1cm}
\caption{Reduced etch rate ($p-1$) \emph{vs.} REL in CR39 for strong and soft
  etching.}\label{calibs}
\end{figure}
The CR39 and Makrofol nuclear track detectors were calibrated with 158 A GeV 
In$^{49+}$ and Pb$^{82+}$ beams at 
the CERN SPS and 1 A GeV Fe$^{26+}$ at the AGS synchroton of the
Brookhaven National Laboratory (BNL). The calibration layout was 
a standard one with a fragmentation target and CR39 (plus Makrofol) NTDs 
in front and behind the target (Cecchini, 2008).
The detector sheets behind the target detected both primary ions and nuclear 
fragments.
 
We recall that  the formation of etch-pit cones (``tracks'') in NTDs is 
regulated by the bulk etching rate, $v_{B}$,  and the track etching 
rate, $v_{T}$, \emph{i.e.} the velocities at which the undamaged and 
damaged materials (along the particle trajectory), are etched out.  
Etch-pit cones are formed if $v_T > v_B$. The response of the CR39
detector is measured by the etching rate ratio $p=v_T / v_B$. 
\begin{figure}[h!]
{\centering\resizebox*{6.5cm}{6.5cm}{\includegraphics{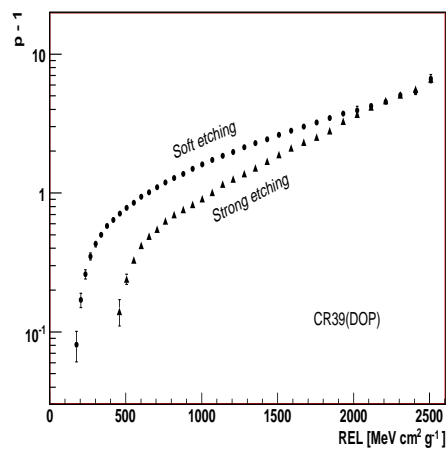}}\par}
\vspace{-0.2cm}
\caption{Reduced etch rate ($p-1$) \emph{vs.} REL for strong and soft
  etching, for CR39(DOP).}\label{calibs2}
\end{figure}
In the calibration procedure, the track's area of each fragment and survival
beam ions are measured, and the $Z$ of each resolved peak is
identified via the average measured base areas.

For each calibration peak the $Z/ \beta$ is obtained 
and the reduced etch rate $(p-1)$ is computed. The Restricted Energy Loss 
(REL) due to ionization and nuclear scattering is evaluated, thus
arriving to the calibration curve $(p-1)$ \emph{vs.} REL. 
Fig. \ref{calibs} and \ref{calibs2} shows the cases for both strong
and soft etching conditions of the CR39 and CR39 with DOP,
respectively. The detection thresholds are listed in table \ref{tab1}.
\vspace{-0.2cm}
\begin{table}[h!]
  \centering
  \begin{tabular}{|c|c|c|c|}
    \hline
	{} & detector type & $Z/\beta$ &$REL$\\
	&  &  & [MeVcm$^2$/g]\\ 
	\hline
	\multirow{3}{*}{strong} & CR39 &14 & 200\\ 
	    {} & CR39 DOP & 21 & 460\\
		{} &  Makrofol & 50 & 2500\\
		\hline
		\multirow{2}{*}{soft} & CR39 & 7 & 50\\
		    {} & CR39 DOP & 13 & 170\\
		    \hline
  \end{tabular} 
    \caption{Detection thresholds $z/\beta$ and restricted energy loss
      REL [MeVcm$^2$/g] for CR39, CR39 DOP, ans Makrofol, respectively.}\label{tab1} 
\end{table}
\vspace{-0.2cm}

\begin{figure}
\begin{center}
\resizebox{6.5cm}{!}{\includegraphics{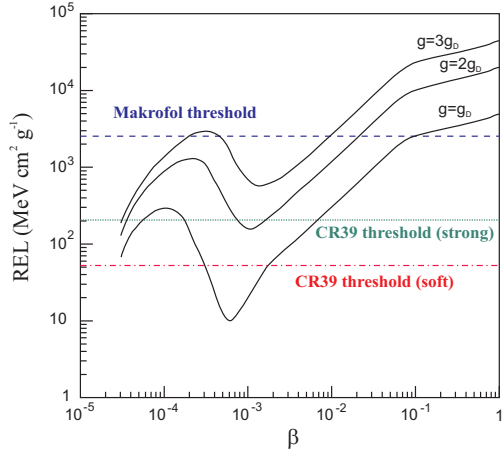}}
\vspace{-0.2cm}
\caption{REL vs beta for magnetic monopoles with $g=g_D,~2g_D,~3g_D$.
The dashed lines represent the CR39 thresholds in soft and strong 
etching conditions and the Makrofol threshold.}  
\label{fig:rel-beta}
\end{center}
\end{figure}

For magnetic monopoles with $g=g_D,~2g_D,~3g_D$ we computed the REL as
a function of $\beta$ taking into account electronic and nuclear energy
losses, see Fig. \ref{fig:rel-beta} (Derkaoui, 1999).

With the used etching conditions, the CR39 allows the detection of $(i)$ 
MMs with $g=g_D$ for $\beta \sim10^{-4}$ and for $\beta > 10^{-2}$; $(ii)$
MMs with $g=2g_D$ for $\beta$ around $10^{-4}$ 
and for $\beta > 4 \times 10^{-3}$; $(iii)$ the whole $\beta$-range of 
$4 \cdot 10^{-5} < \beta < 1$ is accessible for MMs with $g > 2 g_D$ and 
for dyons. 

For the Makrofol polycarbonate the detection threshold is at $Z/\beta \sim50$
and REL $\sim2.5$ GeV cm$^2$ g$^{-1}$ (Balestra, 2007);
 for this reason 
the use of Makrofol is restricted to  the search for fast MMs. 

\begin{figure}[!h]
{\centering\resizebox*{6cm}{7.6cm}{\includegraphics{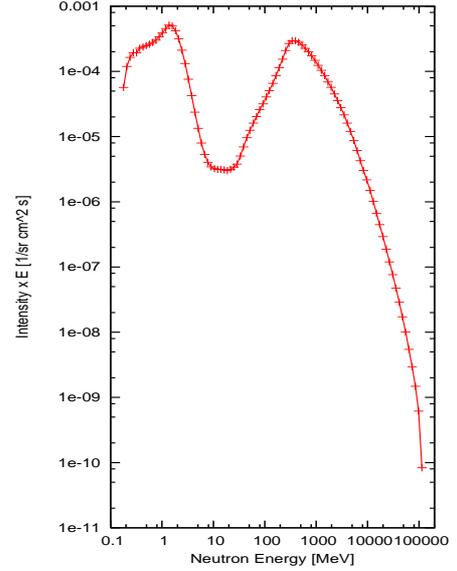}}\par}
\vspace{-0.3cm}
\caption{Wide range spectrometer: experimental neutron spectrum
  measured at Chacaltaya laboratory (5230 m a.s.l., $16^{\circ}$S
  $68^{\circ}$W), Zanini et al. (Zanini, 2001).
}\label{spectrumZa} 
\end{figure}

{\bf{Neutron induced background:}} A statistical study of the
background tracks in the top-most CR39 
layer, generated only by neutrons was made using Monte Carlo (MC)
simulations. The neutron spectrum measured at the SLIM site is shown in
Fig.\ref{spectrumZa}, plotted as the intensity times the energy of the
neutrons \emph{vs.} their energy (Zanini, 2001),
 was used in the simulations.   
The CR39 polymer was defined by its chemical formula, its density and
its post-etched (strong etching conditions) dimensions (Medinaceli, 2008). 

\begin{figure}[htb]
{\centering\resizebox*{6.2cm}{6cm}{\includegraphics{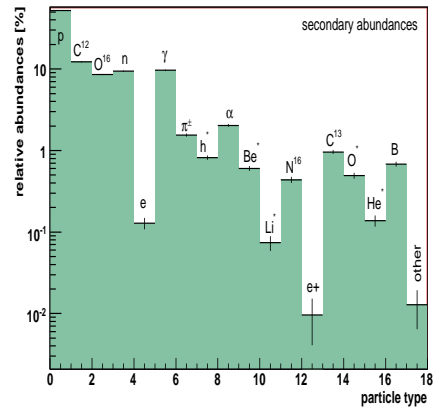}}\par}
\vspace{-0.2cm}
\caption{Secondary production inside CR39 (1000 $\mu$m, post
  etched thick). The relative abundances (ratio of the number of secondaries
  of one specie and the total number of secondaries.) are expresed in
  percentage. Statistical errors are indicated with vertical
  bars.}\label{secondAbundMC}
\end{figure}

For the exposure time, the total number of neutrons impinging on
the detector was calculated, and the relative abundances of secondary
particles generated inside the CR39 was
obtained. Fig. \ref{secondAbundMC} shows the relative abundance of
each kind of secondary particle, obtained as the ratio of the number
of particles of one specie to the total number of secondaries. The
most abundant particles produced are protons, with a relative
abundance of the 52.4\%, then $^{12}C$ isotopes (12.5\%),
$\gamma$-particles (9.5\%), secondary neutrons (9.3\%), $^{16}O$
isotopes (8.5\%), and $\alpha$ particles ($\sim$2\%). The other
species quoted on the plot contribute for less than $\sim$1\%. The
most important mechanism for neutron interaction is 
elastic scattering with the constituent elements of CR39 (H, C,
O). Inelastic scattering is the second most frequent process that
neutrons undergo; 
by-products of this mechanism are $\gamma$ and $\alpha$ particles.
A fraction of the 89\% of the secondary protons have a REL
distribution with values over the detector threshold (considering the
$REL_{th}$ obtained for strong etching). All the $^{12}C$,
$^{16}O$ ions and $\alpha$-particles have REL values over the
$REL_{th}$ value; this implies that all such particles are detectable
by the SLIM detectors, neutrons mainly contribute to the surface
background found on the top-most layer of CR39.\\ 
{\bf{Analysis:}}
The analysis of a SLIM module started by etching the uppermost CR39 sheet 
using strong conditions in order to reduce the CR39 thickness from 
1.4 mm to $\sim0.9$ mm. After the strong etching, the CR39 sheet was scanned 
twice, with a stereo microscope, by different operators, 
 with a 3$\times$ magnification optical lens, looking for any possible 
correspondence of etch pits on the two opposite surfaces. The measured 
single scan efficiency was about 99\%; thus the double scan guarantees an 
efficiency of $\sim100\%$ for finding a possible signal. 

\begin{figure}[htb]
{\centering\resizebox*{5cm}{5cm}{\includegraphics{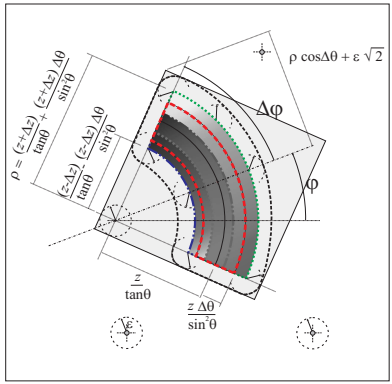}}\par}
\caption{``Confidence'' area in which a possible candidate
  track located on the top layer will be searched for in the lower
  layer of the same module.}\label{regions2}
\end{figure}

Further observation of a ``suspicious correspondence'' was made with an 
optical $20 \div 40$$\times$ stereo microscope and classified either as a 
defect or a candidate track. This latter was then examined by an optical 
microscope with $6.3_{ob} \times 25_{oc}$ magnification and the axes of the 
base-cone ellipses in the front and back sides were measured. A track
was defined as  a ``candidate'' if the computed $p$ and incident angle
$\theta$ on the front and back sides were equal to within 20\%. For  
each candidate the azimuth angle $\varphi$ and its 
position $P$ referred to the fiducial marks were also determined. 
The uncertainties $\Delta \theta$, $\Delta \varphi$, $\Delta z$ (the
error associated to the depth of the lower-most layer of CR39) and
$\Delta P$ defined a ``coincidence'' area ($< 0.5$ cm$^2$) around the
candidate expected position in the other layers, as shown in
Fig. \ref{regions2}. In this case the lowermost CR39 layer was etched
in soft etching conditions, and an accurate scan under an optical
microscope with high magnification (500$\times$ or 1000$\times$) was  
performed in a square region around the candidate expected position, 
which included  the ``coincidence'' area. 
If a two-fold coincidence was detected, the CR39 middle layer was also 
analyzed. 

\section{Results}\label{sec:results}
Since no candidates were found, the 90\% C.L. upper limit for a downgoing 
flux of IMMs and for dyons was computed as 
\begin{center}
\begin{equation}
\phi=\frac{2.3}{(S\Omega) \cdot \Delta t \cdot \epsilon} \nonumber
\end{equation}
\end{center}
where $\Delta t$ is the mean exposure time (4.22 y), $S\Omega$ is the total acceptance,
$\epsilon$ is the scanning efficiency estimated to be $\sim1$. 
  
\begin{figure}[!h]
\begin{center}
  {\centering\resizebox*{10cm}{7cm}{\includegraphics{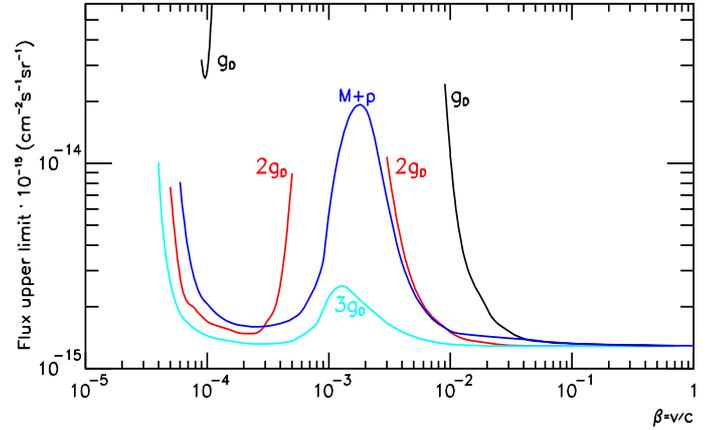}\par}}
\end{center}
\vspace{-0.6cm}
\caption{90\% C.L. upper limits for a downgoing
flux of IMMs with $g=g_D,~2g_D,~3g_D$ and for dyons (M+p , $g = g_D$) 
plotted vs $\beta$ (for strong etching). The poor limits at $\beta
\sim10^{-3}$ arise because the REL is below the threshold (for $g_D$
and $2g_D$) or slightly above the threshold (for $3g_D$ and
dyons).}\label{SLIMlimit}
\end{figure}

The global 90\% C.L. upper limits for the flux of downgoing IMMs and dyons 
with velocities $\beta > 4 \cdot 10^{-5}$ were computed, the complete
dependence with $\beta$ is shown in Fig. \ref{SLIMlimit}. The flux
limit for $\beta > 0.03$ is $\sim1.3 \cdot 10^{-15}$ cm$^{-2}$ s$^{-1}$ sr$^{-1}$. 
\vspace{-0.9cm}
\vspace{-0.4cm}
\section*{Acknowledgements}
We acknowledge the collaboration of E. Bottazzi, L. Degli Esposti and
G. Grandi of INFN Bologna and the technical staff of the Chacaltaya
Laboratory. We thank INFN and ICTP for providing grants for
non-italian citizens.


\begin{thebibliography}{99}
Ambrosio M. et al., 2002. Eur. Phys. J C25,
511. Ambrosio M. et al., 2002. Eur. Phys. J C26, 163.\\ 
Dirac P.A.M., 1931. Proc.~R.~Soc. London 133,p60;
 Phys. Rev. 74(1948)817.\\
Bakari D. et al., 2000. hep-ex/0003028;
Cecchini S. et al., 2001. Il Nuovo Cimento 24C,639.\\
Balestra S. et al., 2008. Eur. Phys. J. C55, p57-63.\\
Balestra S. et al., 2007. Nucl. Instrum. Meth. B254, 254;
Manzoor S. et al., 2005. Radiat. Meas. 40, 433; Nucl. Phys. 
B Proc. Suppl. 172, (2007) 296;
Giacomelli G. et al., 1998. Nucl. Instrum. Meth. A411, 41.\\
Cecchini S. et al., 2001. Radiat. Meas. 34, 55.\\
Cecchini S. et al., 2008. Nucl.Phys. A807, 206.\\
Coleman S., 1985. Nucl. Phys.B262, 293;
Kusenko A. and Shaposhnikov A., 1998. Phys. Lett. B418, 46.\\
Derkaoui J. et al., 1999. Astrop. Phys.10, 339.\\
Giacomelli G. et al., 2000. hep-ex/0005041.\\
Giacomelli G., Manzoor S., Medinaceli E., Patrizii L.,
2007. arXiv:hep-ex/0720502v2.\\
Medinaceli E., 2008. Ph.D. Thesis, Univ. of Bologna.\\
Schraube H. et al., 1999. Rad. Prot. Dos. 84, 309;
Zanini A. et al., 2001. Il Nuovo Cim.24C, 691.\\ 
Sahnoun Z., 2008. arXiv:0805.1797v1; Sahnoun Z. these proceedings.\\
Togo V. and I. Traore, these proceedings; Balestra S. et al.,
2007. NIM B254, 254.\\ 
Witten E., 1984. Phys. Rev. D30, 272;
De Rujula A. and Glashow S., 1984. Nature 31, 272.\\
\end{thebibliography}
\end{document}